\documentstyle[mprocl,psfig,epsf,subfigure]{article}

\renewcommand{\thesubfigure}{\thefigure.\arabic{subfigure}}
\makeatletter
  \renewcommand{\@thesubfigure}{\thesubfigure:\space}
  \renewcommand{\p@subfigure}{}
\makeatother

\title{UNIVERSAL MULTIFRACTAL PROPERTIES OF THE SMALL SCALE INTERMITTENCY IN
ANISOTROPIC AND INHOMOGENEOUS TURBULENCE}

\author{M. ALBER}
\address{Dept. of Medical Physics,\\ Radiologische Uniklinik, Universit\"at T\"ubingen,
D-72076 T\"ubingen, Germany \\ email msalber@med.uni-tuebingen.de}

\author{S. L\"UCK, C. RENNER, J. PEINKE}
\address{FB 8 Physik, Universit\"at Oldenburg, 
D-26111 Oldenburg, Germany \\ email Peinke@uni-oldenburg.de}

\begin{document}
\maketitle

\abstracts{The notion of self-similar energy cascades and multifractality has long 
since been connected with fully developed, homogeneous and isotropic turbulence. 
We introduce a number of amendments to the standard methods for analysing the 
multifractal properties of the energy dissipation field of a turbulent flow.  We 
conjecture that the scaling assumption for the moments of the energy dissipation 
rate is valid within the transition range to dissipation introduced by Castaing 
{\it et al.}( Physica D {\bf 46}, 177 (1990)).  The multifractal spectral functions 
appear to be universal well within the error margins and exhibit some as yet 
undiscussed features. Furthermore, this universality is also present in the neither 
homogeneous nor isotropic flows in the wake very close to a cylinder or the 
off-centre region of a free jet.} 

\section{Introduction}

The concept of an iterated distributive process behind turbulence, commonly 
called the ``energy cascade'', is much older than fractal geometry. It might be 
said that multifractal theory has its roots, partially, in turbulence research 
\cite{Mandelbrot74}. Of course, by being finite the statistics of the energy cascade 
do not have to be multifractal, and even if it may be conceived to be so, the
expedience of a multifractal description is far from obvious. For several decades 
one incentive for turbulence research has been the idea of a universal behaviour in 
small scale intermittency \cite{Frisch95}, but it has not been possible to bring 
this universality to light in the regime of length scales which is accessible to 
a multifractal model. These questions were adressed in pioneering works by Meneveau 
and Sreenivasan \cite{Meneveau87,Sreenivasan91,SreenivasanAntonia97}, and it became 
clear then that multifractal geometry is an issue in the long process of understanding 
turbulence.

This paper is devoted to an extensive examination of experimental data in a multifractal 
framework. We shall not venture to relate our experimental findings to multifractal 
models of turbulence here, however, we vouch for the multifractality of turbulence.

\begin{figure}[htb]%
\begin{center}%
\hspace{-2.0cm}\psfig{file=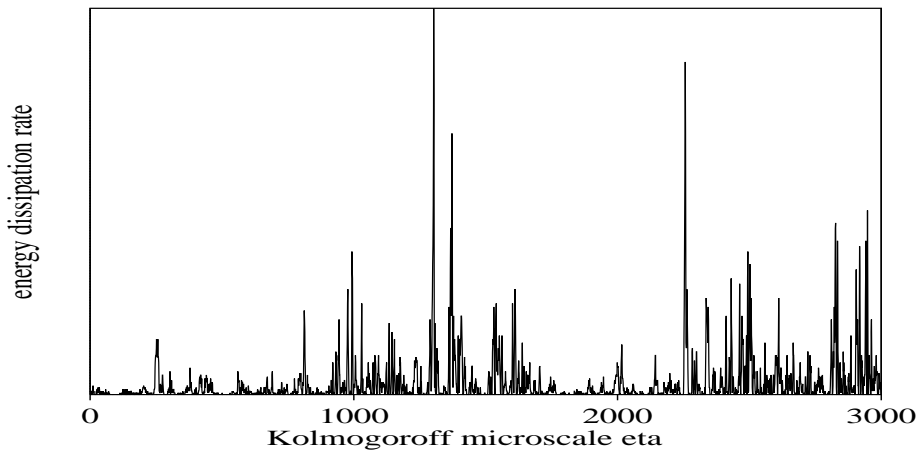,width=14cm,height=6cm}%
\caption{\label{ftshom} \em The energy dissipation time series (equivalent to a 1-D cut along the mean 
flow) for experiment J1. Units on the x-axis are in terms of the Kolmogorov microscale 
$\eta$.} 
\end{center}
\end{figure}

The central quantity of our investigation, as introduced by Obukhov \cite{Obukhov62}  
and Kolmogorov \cite{Kolmogorov62} in 1962, is the energy dissipation averaged over 
a volume of size $r$. In the case of locally homogeneous and isotropic turbulence this 
is approximated by
\begin{equation}
\epsilon_r(x) \quad \propto \quad \frac{1}{r} \int_{x-r/2}^{x+r/2} \left(\frac{\partial u}
{\partial x'} \right)^2 \mbox{d}x' \quad,
\end{equation}
where $u(x)$ denotes the longitudinal velocity component at point $x$ along a 1-D cut 
through the energy dissipation field (Figure \ref{ftshom}).

We define the multifractal measure $\mu_i$ on a grid of size $r$ to be 
$\mu_i = r \epsilon_r(ir)$.
The $q$-th order moment of $\mu$ is defined as :
\begin{equation} \label{partsum}
S_q(r) = \sum_i \mu_i^q \quad \propto \quad r^{\tau(q)} \quad,
\end{equation}
where the sum is taken over all boxes of some $r$-grid and $\tau(q)$
is the multifractal scaling function. 

To derive these quantities from the data, we employ an improved multifractal box counting 
algorithm recently suggested \cite{Alber98}. We sketch this algorithm in section two. 
Section three concerns the details of the experimental data and the particular aspects 
of the selection of the proper scaling range. Section four presents the results obtained 
for locally isotropic flow conditions indicating a universal multifractal behaviour in 
the range defined by the transition from the inertial range to the dissipative limit.
These universal features are also recovered in inhomogeneous and not fully developed 
turbulence, as shown in the concluding section.

\section{An Improved Multifractal Box-Counting Algorithm}

The necessity of taking great care when utilising fixed-size box counting algorithms 
has frequently been expressed. However, here we deem fixed-size algorithms to be the 
method of choice for two reasons: their shortcomings are well understood 
\cite{Caswell87,Provenzale93} and can in some cases be balanced \cite{Alber98,Riedi96}, 
and they are closer to the theoretical concepts of turbulence than their fixed-mass 
counterparts \cite{Badii88} in that they do not mix length scales . We also refrain from 
the direct evaluation of the $f(\alpha)$ curve \cite{Chhabra89}, because we find 
artefacts are easier to trace in the standard approach \cite{Meneveau87,Meneveau91}. 
The information content of both methods is equivalent since they are based on 
essentially the same counting procedure.

We want to outline in brief some of the improvements introduced to the method of moments
\cite{Alber98}. The prime source of errors was identified as the finite size artefacts 
due to ill-fitting grids. To reduce the influence of these ``clipping errors'' we replace
in eq. (\ref{partsum}) the local measure $\mu_i$ by a local average $<\mu>_i$ and define
\begin{eqnarray}
<\mu>_i = \mu_i \prod_{j=1}^n \Theta (\mu(x_i+\delta_j)) \quad & \quad \mbox{for} \quad & q \geq 0 \\
<\mu>_i =  \left( \prod_{j=1}^n \mu(x_i+\delta_j) \right)^{1/n} & \mbox{for} & q < 0 \quad,
\end{eqnarray}
where $\{ \delta_j \}$ is a set of displacements smaller than $r$. 

Secondly, we introduce the concept of wandering intercepts which describe the deviations 
from linear scaling of the partition function (\ref{partsum}) \cite{Alber98,Cutler91,Cutler93}. 
For a given $q$ we define
\begin{equation} \label{wnd}
C(q,r) = \log S_q(r) - \hat{\tau}(q) \log r \quad ,
\end{equation}
where $\hat{\tau}(q)$ is an estimate of $\tau(q)$, e.g. the result of a least squares 
line fit to $\log S_q(\log r)$. The $C(q,r)$ can be thought of as a ``lacunarity function''.
Note that the concept of Extended Self-Similarity introduced by Benzi \cite{Benzi95}
is a special case of eq. (\ref{wnd}). Our development relies on the assumption that the 
$C(q,r)$ does not behave too erratically in $q$ and can ideally be factorised into functions 
of $r$ and $q$ only. Although we shall not make detailed assumptions about the $C(q,r)$ 
in general, a lack of coherence with respect to $q$ at a point $q_0$ betrays the presence
of a phase transitional behaviour of the multifractal \cite{Alber98}. 
For these purposes we employ plots of $C(q_1,r_i)$ against $C(q_2,r_i)$ for 
$q_1 < q_0 <  q_2$ and expect that a lack of correlation between the fluctuations 
in $r$ of the wandering intercepts appears as a scattered plot.

The greatest obstacle in the multifractal analysis of turbulence data (as with any other 
physical data) is the determination of the proper scaling range. For our findings it is
characteristic that the multifractal scaling assumption applies only
to the transitional range to dissipation, which can loosely be defined by the 
lengthscale $\eta$ where the local Reynolds number becomes of the order of unity:
$ \frac{\bar{u} \eta}{\nu} = 1$. 
In most cases, the lower bound is dictated by the detector resolution, which usually 
exceeds $\eta$ quite considerably for turbulent flows of a reasonable Reynolds number.  
The upper bound on the scaling range is much more indistinct, since typically  one can 
only discern a sweeping transition to a power law behaviour of $S_q(r)$, which in 
turn sometimes extends over less than one order of magnitude. Although it is tempting to
expand the scaling range both to acquire a better statistical basis and to extend the 
validity of the analysis, we handle the problem quite restrictively. Since the essence 
of numerical multifractal analysis is extracting to a high precision minor differences 
in interrelated scaling laws, we deem our approach justified.

\section{The Statistical Properties of Experimental Data}

The turbulence velocity data were obtained in three series of experiments of flows
in air. Series J was a free jet through a nozzle of $d=8$ mm diameter and an output 
velocity of 40 m/s. The estimated  $R_{\lambda}$ in the centre was 210. The jet was 
scanned axially and radially at an axial distance of $60d = 48$ cm. The time series
had a size of $1.25 \times 10^7$ data points. Series CF was a wind tunnel experiment 
with a cylinder of diameter $D=5$ cm. The turbulent wake was examined at a distance 
of $32d = 160$ cm for a number of velocities, yielding $R_{\lambda}$ from 200 to 520. 
The size of the time series was $1.25 \times 10^6$. Series CV consisted of a 
variation of the location of the probe in the wake field of the cylinder at 21m/s 
velocity, ranging from $4d = 20 \mbox{cm}$ to $40d = 200 \mbox{cm}$. Again, $1.25 \times 10^7$ 
points were sampled. The equipment was a DANTEC anemometer StreamLine 90CN10 with
single wire probes 55P01 (sensitive length 1.25mm) for the jet measurements and
x-wire probes 55P61 (sensitive length 1mm) for the cylinder measurements. The use
of the latter probe facilitated a more accurate measurement of the longitudinal
velocity component.

\begin{figure}[htb]%
\hspace{1.0cm} \psfig{file=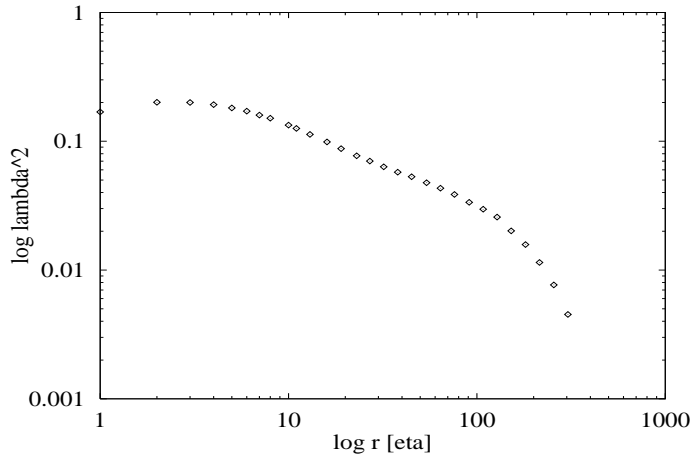,width=10cm,height=6cm}%
\caption{\label{flam} \em A plot of $\log \Lambda^2$ against $\log (r /\eta)$ for experiment J1.
The transitional range corresponds to the  linear part of the curve abutting the 
maximum.}
\end{figure}

An extensive analysis was performed for all time series. This included the 
calculation of characteristic mean-field quantities (see table 1) and the generation 
of probability density functions (PDFs) of velocity increments. The changing form 
of the PDFs as a function of the spatial resolution $r$ was quantified by a parameter 
$\Lambda^2(r)$ according to Castaing \cite{Castaing90}. The power law behaviour of 
the resulting function $\Lambda^2(r)$ locates the transitional range between the 
classical inertial range and the dissipative limit \cite{Chabaud94} (see figure 
\ref{flam}). In the following we restrict our investigations of the scaling 
behaviour of $<\epsilon_r^q>$, respectively $S_q(r)$, to this transitional range. 
Critical evidence for the coincidence of these scaling ranges was deduced from our 
experimental data. The chosen ranges for each example are given in tables 1 and 2.  

It is one of the traits of box-counting, that it delivers a result under almost all
circumstances. However, it is sometimes open to arbitrariness to find out which results 
comprise significant information about the multifractal. As it turned out, statistical
resolution is a severe issue for moments as small as 3 for about 1 million data points,
and 10 million points improve the situation only marginally. It has been shown 
\cite{Meneveau91}, that $f(\alpha)$ will assume negative values due to the 1-D 
nature of the data set as opposed to the three dimensional energy field,
which infers that the main contribution to a moment that corresponds to a negative
$f(\alpha)$ comes from very few singularities. This being the case, it comes as no
surprise to find that the correlation plots for $q_1, q_2 >5$ are extremely well
correlated. In general, we attribute little credibility to any moment larger than 4.
For negative $q$, noise and digitalisation artefacts put an end to the reliability
of the results at sometimes quite moderate values of $q \approx -1$. A special case are
the inhomogeneous examples, where due to the background of the intertwined laminar 
flow no multifractal fine structure can be resolved with negative moments.

\begin{figure}[hbt]%
\label{tqfa}%
\centering%
\subfigure[\em The Experiment J1 serves as a standard against which the others can be compared
since the detector resolution is close to $\eta$. As it turns out, resolution has an impact
both on positive and negative moments. This figure shows the $\tau(q)$ curve of this experiment]
{\label{figreftq} \psfig{file=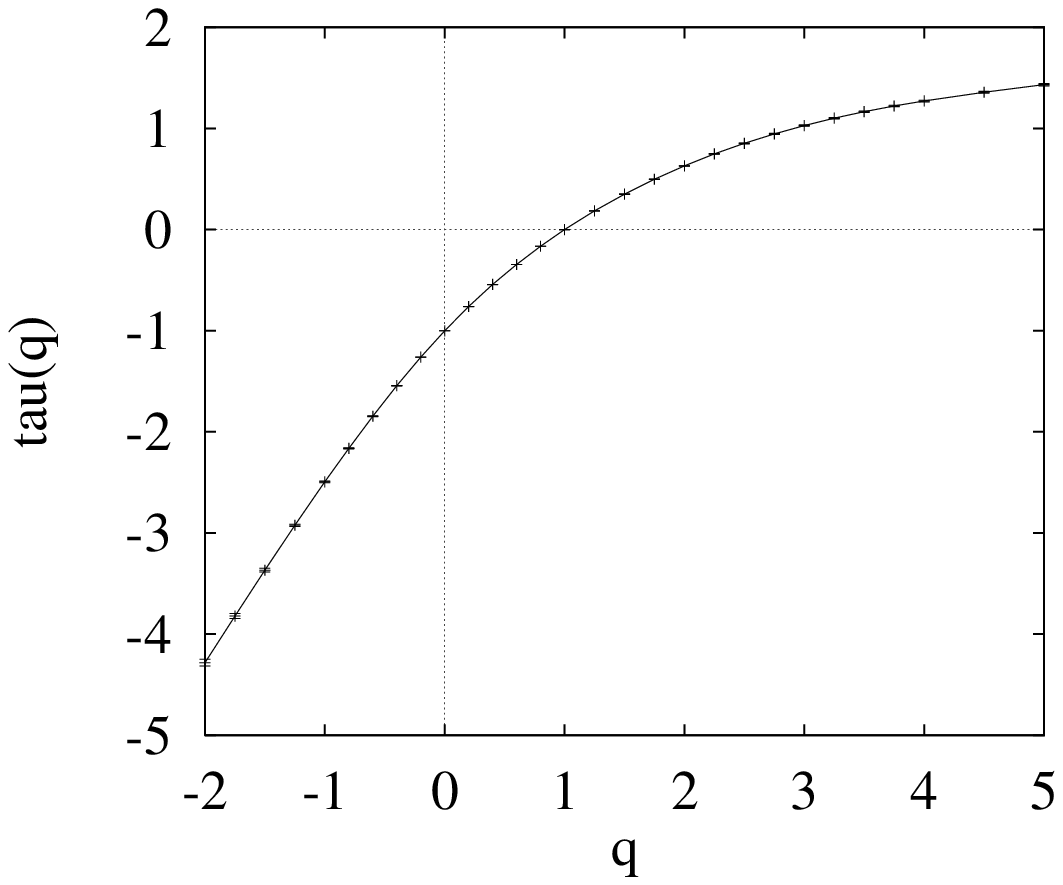,width=6.3cm}}%
\subfigure[\em The $f(\alpha)$ curve corresponding to experiment J1, see also figure 
\ref{figreftq}. This curve was obtained by a Legendre transform of $\tau(q)$. Errors
are amplified due to the great leverage of $q$.]
{\label{figreffa} \psfig{file=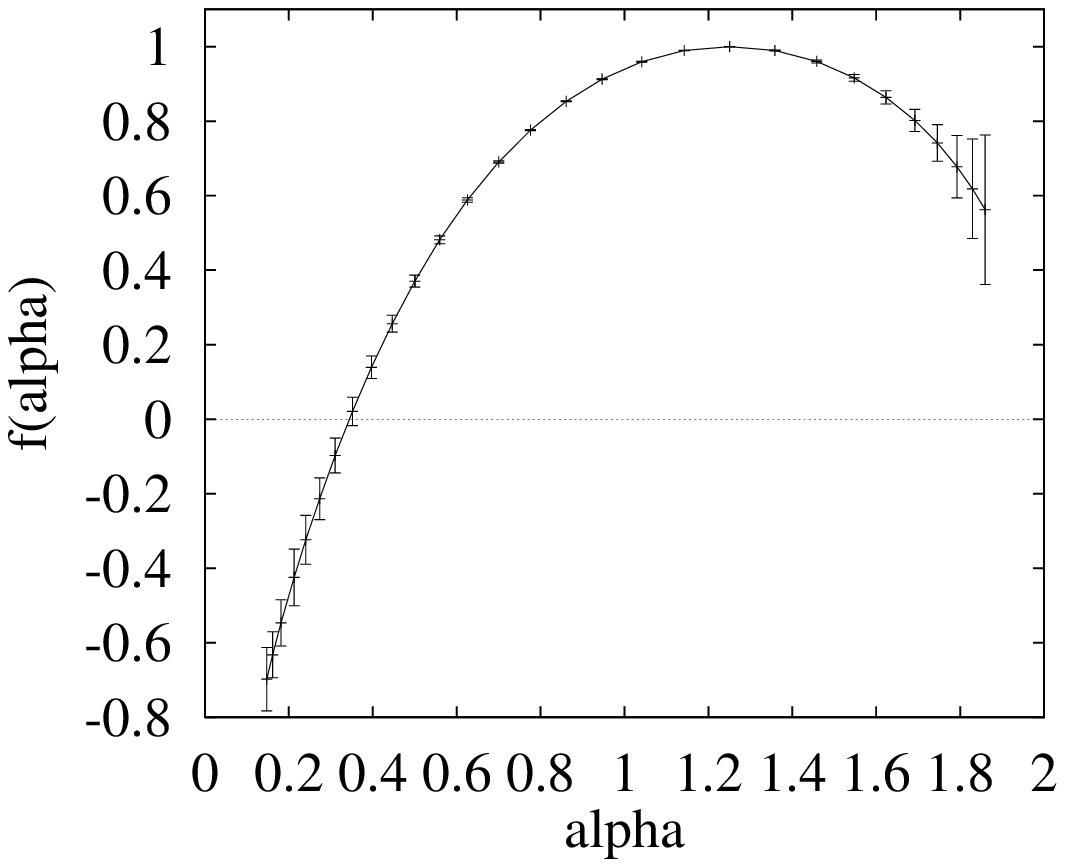,width=6.3cm}}%
\end{figure}
\addtocounter{figure}{1}

\section{Multifractal Spectra in the Locally Isotropic Case}

In the following, four examples of classical fully developed turbulence are
investigated in the centre of the flows, where local isotropy is 
known to hold for small scales. The experiments were chosen to cover a range 
of $R_{\lambda}$ from 190 to 520.
\mbox{Table 1} gives the relevant details. An example of an energy time series
is given in \mbox{figure \ref{ftshom}.} 

\begin{table}[h]
\begin{center}  
\begin{tabular}{|l|c|c|c|c|} \hline 
flow & free jet & free jet & wake  & wake  \\
 & & & of a cylinder & of a cylinder \\ \hline \hline
 reference & J1 & J2 & CF1 & CF2 \\ \hline
position of probe & 125 diam. & 60 diam. & 32 diam. & 32 diam. \\ \hline
number of points & $1.25\times 10^7$ & $1.25\times 10^7$ & $1.25\times 10^6$  & $1.25 \times 10^6$ \\ 
\hline
mean velocity & 2.26 m/s & 4.2 m/s & 22.2 m/s & 4.6 m/s \\ \hline
sampling rate & 8 kHz & 50 kHz & 120 kHz & 20 kHz \\ \hline
Kolmogorov & 0.25 mm & 0.1 mm & 0.07 mm & 0.22 mm \\
microscale $\eta$ & & & & \\ \hline
Taylor microscale $\lambda$ & 27 $\eta$ & 28 $\eta$ & 44 $\eta$ & 29 $\eta$ \\ 
$\lambda = \frac{\langle(u-\bar{u})^2\rangle}{\langle (\partial_x(u-\bar{u}))^2 \rangle}$ & & & & \\ \hline
Taylor-Reynolds number& 190 & 210 & 520 & 220 \\
 $R_{\lambda} = \frac{\lambda \sqrt{\langle (u - \bar{u})^2 \rangle}}{\nu}$ & & & & \\ \hline
transitional range & 4 .. 113 $\eta$ & 8 .. 83 $\eta$ & 26 .. 79 $\eta$ & 8 .. 85 $\eta$ \\ \hline
detector resolution & 4 $\eta$ & 10 $\eta$ & 14 $\eta$ & 5 $\eta$ \\ \hline
permissible moments & -2 .. 5 & -0.2 .. 4 & -0.6 ..  4 & -4 .. 4 \\ \hline   
\end{tabular}
\end{center}
\vspace{.3cm}
Table 1 {\em Summary of experimental conditions and evaluation parameters 
for the locally isotropic flows.}
\end{table}

In the context of this paper, we want to establish experiment J1 as the standard for a 
simple reason: the detector resolution and $\eta$ are quite close, so that one can
assume that the fine structure of the original velocity signal is well resolved. In fact,
a direct comparison with experiment J2 shows that this point is crucial for the
calculation of extreme moments. Figures \ref{figreftq} and \ref{figreffa} 
show the $\tau(q)$ and $f(\alpha)$ curve for the transitional range.

\begin{figure}[htp]%
\hspace{1cm} \psfig{file=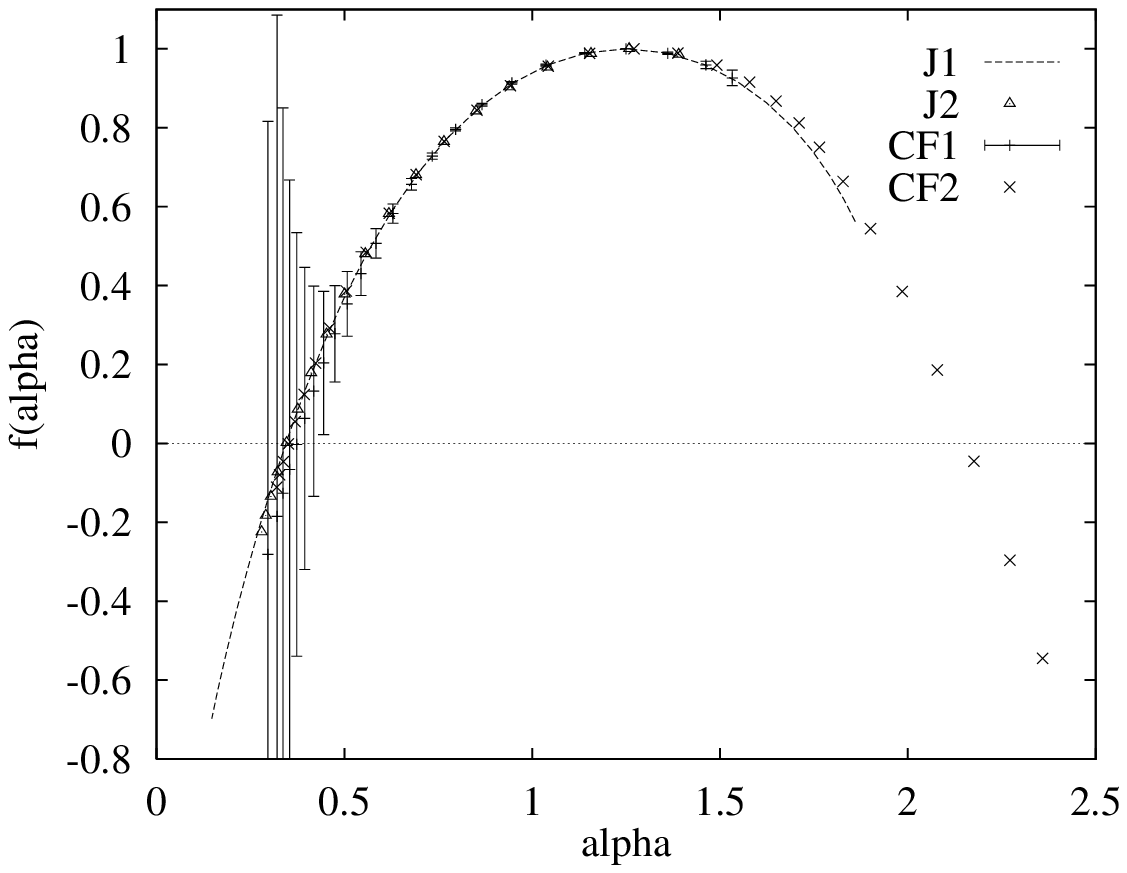,width=11cm,height=6.8cm}%
\caption{\label{fhtb} \em The superposition of the $f(\alpha)$ curves of all locally isotropic
experiments (J1, J2, CF1, CF2) are in reasonable agreement. The solid line is the standard J1, 
the errorbars given for CF1 are representative for all experiments with 'poor' statistics.}
\end{figure}

Figure \ref{fhtb} shows
a superposition of the respective $f(\alpha)$ spectra for all experiments of table 1.
The errorbars are significantly larger than the deviations between the curves. Given 
these results, we have to assume universality of the spectra of singularities.

Although it is notoriously difficult to establish negative values of $f$ reliably,
we have confidence in the results we have given. A self consistent analysis \cite{Alber98}
of the $\tau(q)$ curve showed that statistics are still sound for $q \approx 3$, 
whereas the intersection with the $\alpha$-axis occurs for $q=3.2$ c.f. figure \ref{tqfa}.

There is a couple of interesting, but less obvious results. Firstly, from correlation
plots of $C(q,r)$ one is led to deduce the presence of a phase transition at $q_0 = 1$, which
does not express itself in a kink in $f(\alpha)$ since both tangents necessarily
have slope 1. We interpret this as the presence of two phases: the highly intermittent
or turbulent phase which dominates the moments for $q>1$, and the ''laminar'' phase
which does so by the less pronounced singularities for moments with $q<1$. 

Secondly, if one sets a threshold for the analysis of the measure, one finds that 
for all (very small) thresholds the laminar branch of the multifractal spectrum
 disappears. The conclusion is, that all energy dissipation must be concentrated
on the set which corresponds to the H\"older exponent $\alpha(q=1)$. Whilst this
is necessarily so for mathematical multifractals, it is remarkable for a physical,
hence finite one. In other words, the steepest singularities feed on most of the 
energy and dissipation attracts further energy to dissipate.

\pagebreak
\section{Multifractal Spectra in the Inhomogeneous Case}

\begin{figure}[htbp]%
\label{finhom}%
\centering%
\subfigure[\em The energy time series of experiment J3 in the off-centre region of a free jet.
The width was chosen to match figure \ref{ftshom} in terms of $\eta$. Laminar phases
are mixed with turbulent bursts which were transported out of the central region of the jet.]
{\psfig{file=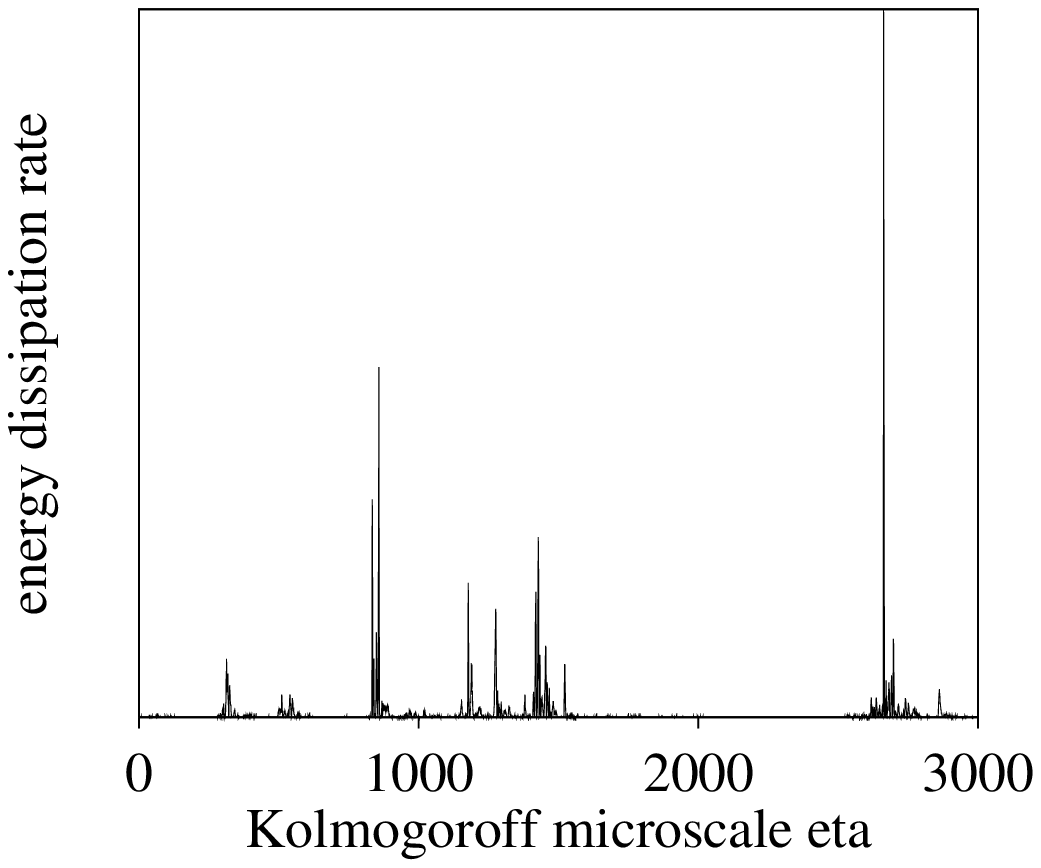,width=6.2cm}}%
\subfigure[\em The energy time series of experiment CV1 in the wake close behind the cylinder.]
{\psfig{file=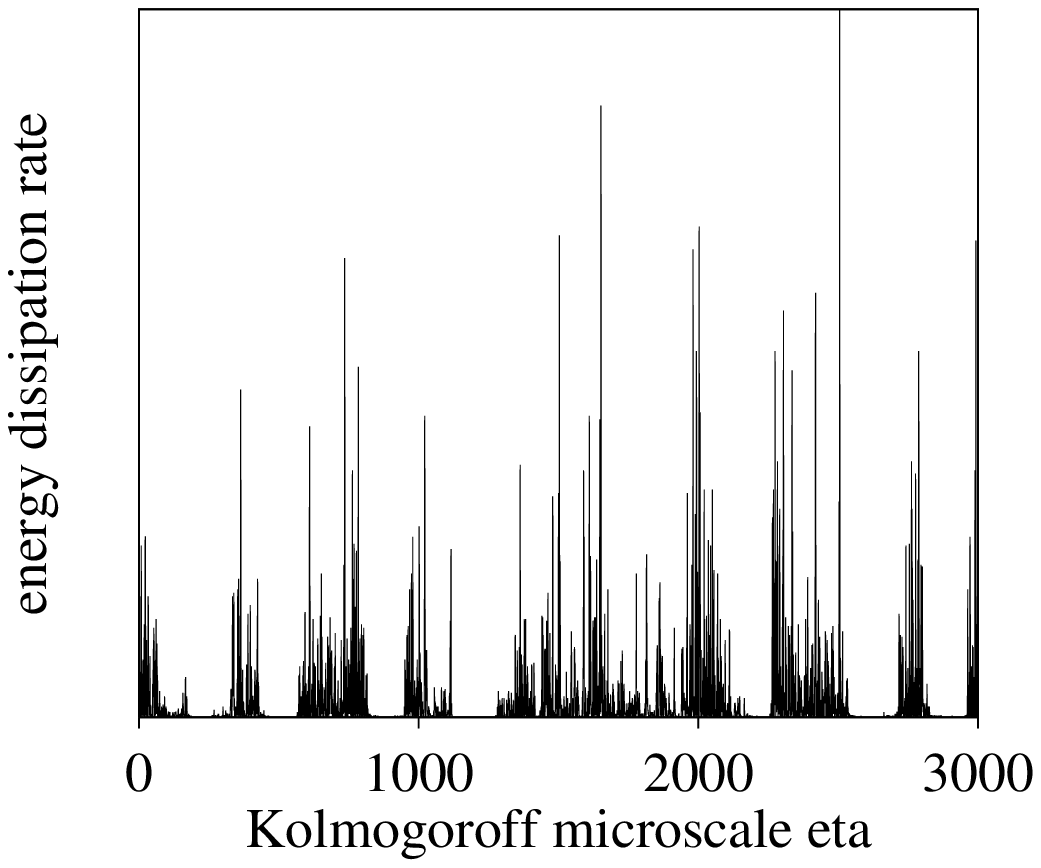,width=6.2cm}}%
\end{figure}

Inherent to most theories pertinent to small scale turbulence is the assumption of isotropy
and homogeneity of the turbulent velocity field. A question of practical importance
is if similar properties can be found where these conditions are violated.
To approach this we investigated a free jet with an radial offset of 9 cm in 48 cm 
axial distance
from the nozzle, i.e. in the outer layers where laminar phases are mixed in.
As a second inhomogeneous situation the wake of a cylinder was examined in 4 diameters 
distance, i.e. under conditions of not fully developed turbulence.
\mbox{Figure \ref{finhom}} shows the energy dissipation field for these cases.
In comparison to figure \ref{ftshom} we clearly see intermittent
burst of turbulent activity in an otherwise smooth flow.

\addtocounter{figure}{1}
 
\begin{table}[hp]
\begin{center}
\begin{tabular}{|l|c|c|} \hline
flow & free jet & wake \\
 & off-centre  & near a cylinder \\ \hline \hline
 reference & J3 & CV1 \\ \hline
position of probe & 60 diam. & 4 diam. \\ 
& 11.25 diam. off-centre & \\ \hline
number of points & $1.25\times 10^7$ & $1.25\times 10^7$ \\ \hline
mean velocity & 0.43 m/s & 21.2 m/s \\ \hline
sampling rate & 6 kHz & 200 kHz \\ \hline
Kolmogorov & 0.17 mm & 0.035 mm \\
microscale $\eta$ & & \\ \hline
Taylor microscale $\lambda$ & 11 $\eta$ & 53 $\eta$ \\ \hline
Taylor-Reynolds & 34 & 720 \\
 number $R_{\lambda}$ & &\\ \hline
transitional range & 4 .. 25 $\eta$ & 18 .. 129 $\eta$ \\ \hline
detector resolution & 6 $\eta$ & 28 $\eta$ \\ \hline
permissible moments & 0 .. 4 & 0 .. 4 \\ \hline   
\end{tabular}
\end{center}
\vspace{.3cm}
Table 2:{\em Summary of experimental conditions and evaluation parameters 
for the inhomogeneous flows.}
\end{table}

The analysis is to some extent different from the homogeneous case. Firstly, most 
quantities like $R_{\lambda}$ and $\eta$ loose their physical meaning.
Secondly, the $\Lambda^2$ plots become distorted by the laminar
phases, so that the analysis is open to some greater degree of arbitrariness than in
the homogeneous case. Table 2 gives the approximate values.

\begin{figure}[htnp]
\hspace{1.5cm} \psfig{file=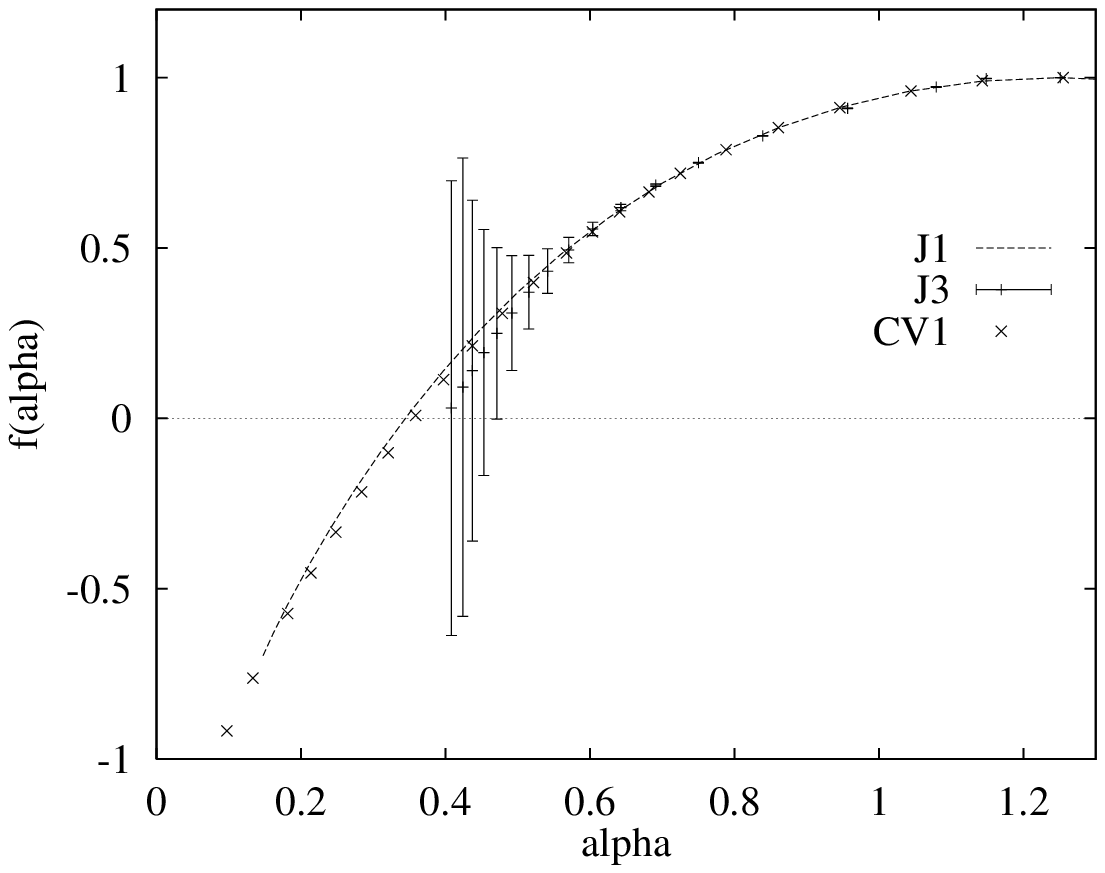,width=10cm}
\caption{\label{fiht} \em In the inhomogeneous case, the right branch of $f(\alpha)$ is not accessible.
The solid line again is the standard J1, errorbars are given for J3. Statistcs are quite 
poor for the jet measurement due to the great dilatation of the turbulent bursts.}
\end{figure}

From the spectra in Figure \ref{fiht} it could be deduced that the small scale
properties of the energy dissipation field are to some degree indepedent of the 
prerequisites of global uniformity and isotropy. We cannot state yet to which extent
these conditions do not hold within a turbulent burst, but we accept the possibility
that the multifractal aspects of the energy dissipaion field are in fact universal.

Notice that the `laminar' branch of the spectrum is not accessible due to the presence
of a laminar background. Thus the right hand part of the spectrum collapses to a point 
at the maximum. The peak at unity indicates that the energy dissipation is still space
filling locally, despite its sparse appearance. 

\section{Conclusion} 

We have reported on a correspondence in small scale energy dissipation spectra
in both locally isotropic and inhomogeneous turbulence. 
The results are plausible if one takes into consideration
that firstly the analysis is restricted to small scales of less than 
100$\eta$, and secondly, by the observation that the inhomogeneous turbulence in the
boundary layer of the free jet can be thought of as a composition of zones of turbulence
transported out of the centre and inserts from the surrounding laminar flow 
\cite{Sreenivasan91}.
Less accessible to intuition are our findings for the non-developed turbulent wake.

To set our findings in context with previous work on small scale turbulence,
we want to stress once more that our work focuses on the transitional regime,
a range of length scales between the inertial range and the dissipative limit. 
Here we expect that the dissipation plays a 
role of increasing importance. Traditionally, scaling behaviour is investigated 
on larger scales in the inertial range, where the dissipation should be of minor 
influence. 
Some recent work has shed new light on the turbulent cascade in the inertial range.
\cite{Frisch95,Greiner97,Nelkin96,Naert97}. Our results suggest that in the neighbouring
regime the statistics of the energy dissipation field $\epsilon_r$ become universally
multifractal.
The challenge remains to match the various regimes in which turbulence is described
to finally arrive at a truly comprehensive view.

\section{References}
\bibliographystyle{unsrt}
\bibliography{diplu}

\end{document}